\documentclass[aip,pop,reprint,superscriptaddress]{revtex4-1}
\usepackage{dcolumn}% Align table columns on decimal point
\usepackage{bm}
\usepackage{amssymb,amsmath,amsthm,epsfig}
\usepackage{sidecap}

\usepackage{natbib}

\def\apj{ApJ.}
\def\solphys{Sol. Phys.}
\def\aap{A\&A}

\def\jgr{J. Geophys. Res.}

\def\ssr{Space Science Rev.}

\DeclareMathOperator\erf{erf} 

\begin{document}

\title{Reconnection at 3D Magnetic Null Points: Effect of Current Sheet Asymmetry}
\author{P. F. Wyper}
\email{app09pfw@sheffield.ac.uk}
\affiliation{School of Mathematics and Statistics, University of Sheffield, S3 7RH, UK}
\author{Rekha Jain}
\email{R.Jain@sheffield.ac.uk}
\affiliation{School of Mathematics and Statistics, University of Sheffield, S3 7RH, UK}

\date{\today}

\begin{abstract}
Asymmetric current sheets are likely to be prevalent in both astrophysical and laboratory plasmas with complex three dimensional (3D) magnetic topologies. This work presents kinematic analytical models for spine and fan reconnection at a symmetric 3D null with asymmetric current sheets. Asymmetric fan reconnection is characterized by an asymmetric reconnection of flux past each spine line and a bulk flow of plasma across the null point. In contrast, asymmetric spine reconnection is inherently equal and opposite in how flux is reconnected across the fan plane.  The higher modes of spine reconnection also include localized wedges of vortical flux transport in each half of the fan. In this situation, two definitions for reconnection rate become appropriate: a local reconnection rate quantifying how much flux is genuinely reconnected across the fan plane and a global rate associated with the net flux driven across each semi-plane. Through a scaling analysis it is shown that when the ohmic dissipation in the layer is assumed to be constant, the increase in the local rate bleeds from the global rate as the sheet deformation is increased. Both models suggest that asymmetry in the current sheet dimensions will have a profound effect on the reconnection rate and manner of flux transport in reconnection involving 3D nulls. 
\end{abstract}


\maketitle

\section{Introduction}
The importance of magnetic nulls to magnetic reconnection has long been recognized. In two dimensions (2D) reconnection only occurs where there is a hyperbolic null (or X-point) in the magnetic field. Nulls of this type form the basis for the classic two dimensional reconnection models of Sweet-Parker \cite{Sweet1958,Parker1957} and Petscheck \cite{Petschek1964}. However, as an increasingly accurate picture of the complex three dimensional (3D) nature of the solar and magnetospheric magnetic fields is developing, the importance of the fully 3D null point is now being more appreciated. 

For instance, during quiet sun periods 3D nulls are found in abundance in the lower solar atmosphere \cite{Regnier2008,Longcope2009}, whereas in active times of the solar cycle they play a role higher up and are believed to be involved in solar flares \cite{Aulanier2000}, magnetic breakout \cite{Lynch2008}, jets \cite{Pariat2009,Liu2011}, flux emergence \cite{Torok2009} and flare brightening \citep{Masson2009,Fletcher2001}. Through \emph{in situ} observations 3D nulls have been confirmed to exist in the earths magnetotail \citep{Xiao2006}, as well as being inferred through global simulations to exist in clusters within the polar cusp regions \citep{Dorelli2007}. In certain 3D laboratory experiments reconnection at 3D nulls also plays an important role \cite{Gray2010}. 

The field structure of the 3D null is somewhat different from the 2D X-point and can be described via a Taylor expansion in the vicinity of the null so that
\[\mathbf{B}=\mathcal{M}\cdot \mathbf{r}\]
where $\mathcal{M}$ is the Jacobian of $\mathbf{B}$ and $\mathbf{r}$ is the position vector $(x,y,z)^{T}$. The simplest linear potential null can be expressed accordingly as 
\[\mathbf{B}=\frac{B_{0}}{L_{0}}(x,\kappa y,-(1+\kappa )z),\]
where $\kappa$ is a dimensionless constant and $B_{0}$ and $L_{0}$ are some typical field strength and length scale respectively \cite{Parnell1996}. The eigenvectors of $\mathcal{M}$ (whose corresponding eigenvalues sum to zero since $\boldsymbol{\nabla} \cdot \mathbf{B}=0$) define the spine and fan such that the two eigenvalues whose real parts have like sign lie in the fan plane with the third directed along the spine line. The fan plane is a separatrix surface and separates two topologically distinct regions. In the special case of $\kappa=0$ the spine expands into the $y$-direction and the null becomes an X-line.

Reconnection occurring within the current layers which form at 2D X-points only takes the form of a one-to-one breaking and rejoining of the magnetic field. However, at fully 3D null points new connections form in a variety of different ways. Twisting motions about the spine/fan lead to the formation of a current sheet aligned to the fan/spine within which \emph{torsional fan/spine reconnection} occurs \cite{Wyper2010, AlHachami2011, PriestPontin2009}. Shearing motions across the spine/fan lead to current sheets forming aligned to the fan/spine within which \emph{fan} and \emph{spine} reconnection occur \cite{PriestTitov1996,LauFinn1990}. Exact incompressible models exist for these modes utilizing current layers of reduced dimensionality \cite{CraigFabling1996,Heerikhuisen2004}. Such solutions are sometimes referred to as reconnective annihilation \cite{PriestForbes2000} since the infinite extent of the current layers means that once field lines thread into a current layer they never leave. Therefore, there is no `other end' of the field line in an ideal region which this field may be `reconnected' to. What field is washed into the current layers (crossing the spine or fan in the process) is instead dissipated through ohmic heating. When the incompressibility assumption is relaxed, however, the current layer which forms is localized around the null (locally spanning both the spine and fan) within which a combination of both spine and fan reconnection occur known as \emph{spine-fan reconnection} \cite{PriestPontin2009}. Conceptually, this combination is similar to the 2D scenario as the magnetic flux crosses both the spine and the fan outside of the non-ideal region, where field lines are frozen to the plasma, so that flux is genuinely `reconnected' as a result.

Typically, investigations of both 2D and 3D null reconnection focus on the symmetric case where flux is fed into and removed from the non-ideal region in a symmetric manner. However, there are many situations where this is not the case including at the Earth's magnetopause \cite{Paschmann2013}, during the occasional CME and solar flare \cite{Murphy2012} and in certain laboratory experiments (e.g. the `pull' and `push' modes of the Magnetic Reconnection Experiment (MRX) \cite{Yamada1997,Murphy2008}). 

In 2D magnetohydrodynamics (MHD) the electric field is perpendicular to the magnetic and velocity fields ($\mathbf{E}\cdot\mathbf{B}=0$) so that the connection change which occurs in the current layers is in a pairwise fashion \cite{Hornig1996} with the electric field at the null giving the absolute rate of reconnection. Asymmetry in the 2D literature typically denotes asymmetric upstream/down densities and magnetic field strengths, with investigations of such asymmetries focusing on how the asymmetry affects the absolute rate of reconnection \cite{Murphy2010,Cassak2007}, for a current sheet of fixed length. Phenomenologically, a measurable result of such asymmetry is the displacement of the flow field stagnation point from the null. 

In 3D, the important quantity for reconnection is the component of the electric field parallel to the magnetic field ($E_{\parallel}$), with the maximum of $\int{E_{\parallel} dl}$ along all field lines threading the non-ideal region giving the measure of the reconnection rate \cite{Hesse1988}. Therefore, of particular importance to the 3D reconnection rate is the dimensions of the non-ideal region and the strength of $E_{\parallel}$ within it. Due to the differing magnetic field geometry, reconnection involving 3D nulls can become asymmetric in one of two ways. Firstly, the null field itself may have inherent asymmetry. That is, the eigenvalues associated with the fan plane are of different values ($\kappa\neq1$). This leads to asymmetric current sheet formation and a reconnection rate which depends upon $\kappa$ \cite{AlHachimi2010,AlHachami2011,Galsgaard2011}. Alternatively, the null field itself may be symmetric ($\kappa=1$) but, through the manner of external driving or local plasma anisotropy, the current sheet that forms at it is not. Recent work by \citet{Wyper2012} and \citet{Wyper2013} has shown that even with an initially symmetric null and a homogeneous plasma such current sheet asymmetries can arise due to transient effects. \citet{Watson1997} are, to date, the only investigation to construct asymmetric analytical current sheet solutions at 3D nulls. They noted in a broad investigation of different fan reconnection solutions that asymmetric hyperbolic solutions of reduced dimensionality were possible to construct but did not pursue this further.

The principle aim of this paper is to develop asymmetric analytical models to investigate the consequences of current sheet asymmetry for reconnection at 3D magnetic null points. Specifically, we will develop kinematic models of the spine and fan reconnection modes with asymmetric current sheets and show that each mode has a distinct and different behavior. The layout of this paper is as follows. In Section \ref{sec:2} we introduce the analytical methodology. Section \ref{sec:3} introduces the model for fan reconnection and Sections \ref{sec:4} and \ref{sec:rrsimpleas} introduce a simple model for spine reconnection and how asymmetry affects the reconnection rate in this case. In Sections \ref{sec:asymgen} and \ref{sec:conclusion} we discuss further the consequences of current sheet asymmetry for the reconnection rate using more complex spine models and conclude our findings.

\section{General Method}
\label{sec:2}
We consider various models which are solutions of the steady state kinematic resistive MHD equations given by
\begin{eqnarray}
\mathbf{E}+\mathbf{v}\times\mathbf{B}&=&\eta \mathbf{J}, \quad \mathbf{\nabla}\times\mathbf{E}=0, \nonumber\\
\mathbf{\nabla}\times \mathbf{B} &=& \mu_{0} \mathbf{J}, \quad \mathbf{\nabla} \cdot \mathbf{B} =0. \label{current}
\end{eqnarray}
In each we start with a linear potential symmetric magnetic null of the form
\begin{equation}
\mathbf{B}_{n}=\frac{B_{0}}{L_{0}}\left(x,y,-2z\right),
\end{equation}
to which some localized perturbation field $\mathbf{B}_{p}$ is added such that the total magnetic field is given by
\[\mathbf{B}=\mathbf{B}_{n}+\mathbf{B}_{p}.\]
A symmetric null is chosen as the background field for these models so that only the effects of asymmetry from the perturbation field is important. The electric field and perpendicular plasma flow are then found using 
\begin{eqnarray}
\frac{d \mathbf{X}(s)}{d s} &=& \mathbf{B}(\mathbf{X}(s)), \quad \Phi = -\int{ \eta \mathbf{J}\cdot \mathbf{B} ds} +\Phi_{0}, \label{Phi}\\
\mathbf{E}&=&-\boldsymbol{\nabla}\Phi, \quad \mathbf{v}_{\perp}=\frac{(\mathbf{E}-\eta\mathbf{J})\times\mathbf{B}}{B^2}, \label{vperp}
\end{eqnarray}
where $s$ is related to the distance along a field line through $ds=dl/|\mathbf{B}|$ and $\boldsymbol{\nabla}\Phi_{0}\cdot\mathbf{B}=0$, so that $\Phi_{0}$ is identified with a global ideal background electric field. Solutions with $\Phi_{0}\neq 0$ are known as \emph{composite solutions} and couple the local non-ideal region to the global field. However, we set $\Phi_{0}=0$ in these models and focus on solutions purely of the non-ideal integral term in Equation \ref{Phi}, known as \emph{pure solutions}, which show how flux is reconnected locally. Composite solutions are deferred for later work. For clarity we denote $\Phi$ as $\Phi_{ni}$ from now on.

\section{Asymmetric Fan Reconnection}
\label{sec:3}
To model asymmetric fan reconnection
\begin{eqnarray}
\mathbf{B}_{p}&=&f(x,z) \;\mathbf{\hat{y}}, \nonumber\\
&=&-\frac{jB_{0}}{L_{0}}z e^{-\frac{z^2}{h^2}-\frac{(zx^2)^{2}}{l^6}} g(z) \;\mathbf{\hat{y}},
\end{eqnarray}
is chosen so that the field perturbation (and therefore the current, $\mathbf{J}$) is localized in $x$ and $z$ and asymmetry can be introduced through the weighting function $g(z)$. The parameters $l, k$ and $j$ control the sheet thickness, width and strength respectively. The field line equations are then given by
\[ x=x_{0} e^{B_{0}s/L_{0}}, \; z=z_{0}e^{-2B_{0}s/L_{0}}, \; Y=Y_{0} e^{B_{0}s/L_{0}}, \]
where $Y_{0}(x_{0},y_{0},z_{0})$ is a constant of integration and
\begin{eqnarray}
Y&=&y-e^{B_{0}s/L_{0}}\int \! e^{-B_{0}s/L_{0}} f(x,z) \, \mathrm{d}s, \nonumber \\
&=& y-\frac{j}{2} z^{-1} e^{-\frac{(zx^{2})^{2}}{l^{6}}} I_{1}(z). \label{Y}
\end{eqnarray}
In general, the function $I_{a}(z)$ is given by
\[ I_{a}(z)=z^{\frac{1}{2}} \int{ z^{\frac{a}{2}} e^{-\frac{z^{2}}{h^{2}}} g(z) dz}, \]
using the fact that $ds=dz/B_{z}$ and that both $zx^2$ and $zY^2$ are independent of $s$. Surfaces of field lines are described by $C_{1}(zx^2)=const.$ and $C_{2}(zY^2)=const.$, where $C_{1}$ and $C_{2}$ are arbitrary functions which are independent of $s$.

Assuming a resistivity localized in the $y$-direction,
\begin{equation}
\eta=\eta_{0}e^{-\frac{\left(zY^{2}\right)^{2}}{l^6}},
\end{equation}
the electric potential can be obtained as
\begin{eqnarray}
\Phi_{ni} &=& \frac{j\eta_{0}B_{0}}{\mu_{0}L_{0}} x \left[ \frac{1}{2}\left(1-\frac{2}{l^{6}}(zx^{2})^{2}\right) I_{-3}\left(z\right) -\frac{1}{h^{2}} I_{1}\left(z\right) \right.\nonumber\\
&\quad&\left. \quad -\;\frac{4}{l^{6}}zx^{2} I_{3}\left(z\right) + \frac{1}{2} K_{-1}(z)\right] e^{-\frac{z^{2}(x^4+Y^4)}{l^6}},  \label{Phi:fan}
\end{eqnarray}
where
\begin{equation*}
K_{a}(z)= z^{\frac{1}{2}} \int{ z^{\frac{a}{2}} e^{-\frac{z^{2}}{h^{2}}} g'(z) dz}
\end{equation*}
and $'$ denotes $d/dz$. $K_{a}(z)$ can also be related to $I_{a}(z)$ using integration by parts, although it is more convenient to leave it in this form.

\begin{figure}
\centering
\includegraphics{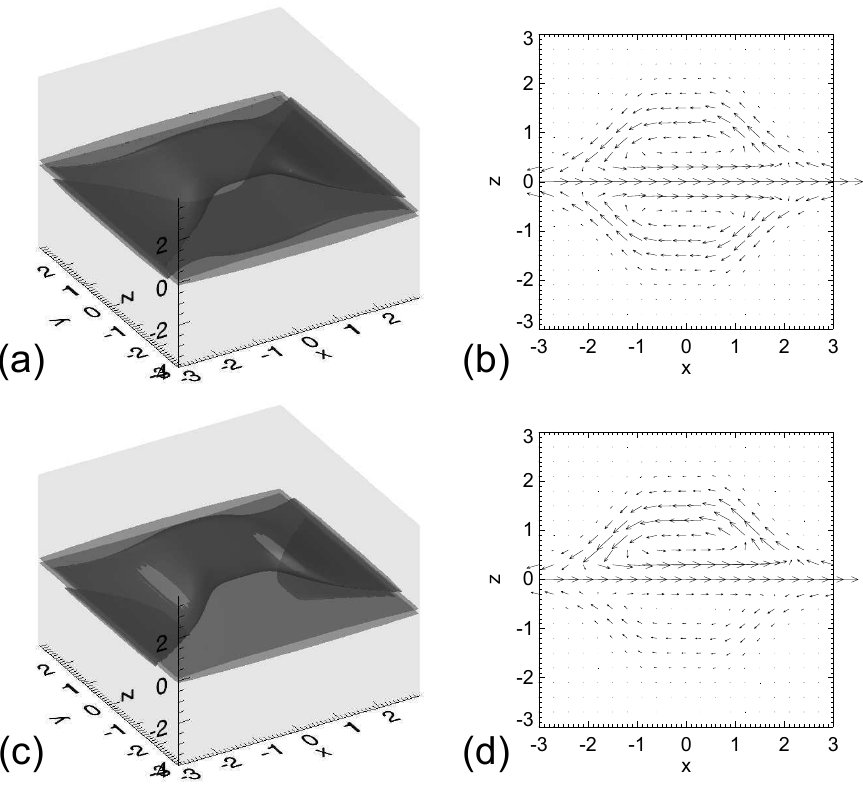}
\caption{(a)-(b) isosurface of $\eta|\mathbf{J}|$ at 25\% of the maximum and the current flow in the $y=0$ plane for the symmetric model. (c)-(d) the equivalent figures for the asymmetric model with $p=0.5$ and $m=0.5$ (see Equation \ref{asymsp-g}). Both have the parameter set $(B_{0},L_{0},\eta_{0},\mu_{0},j,h,l)=(1,1,1,1,2,1,2^{1/3})$.}
\label{fig:fan-etaj}
\end{figure}

\begin{figure}
\centering
\includegraphics{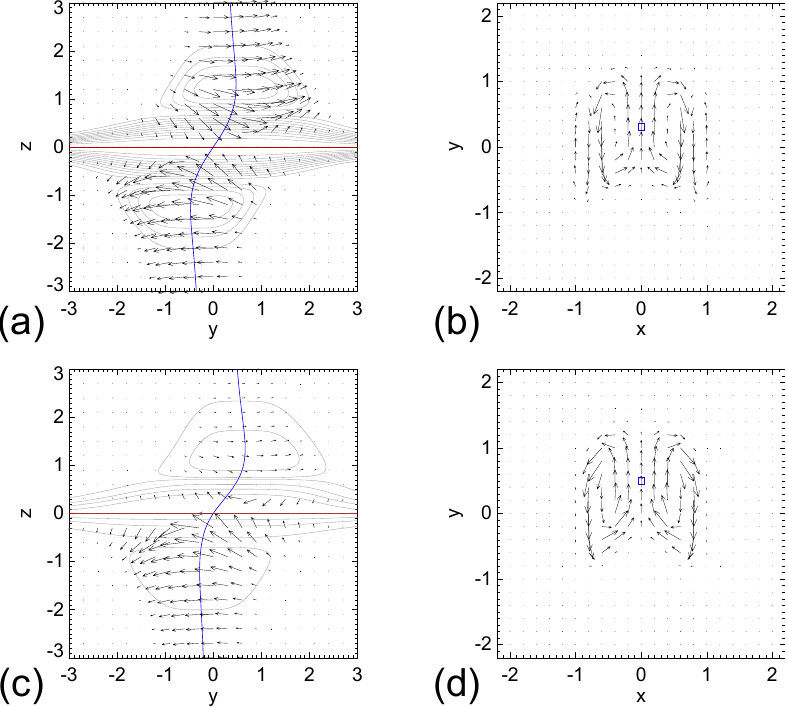}
\caption{(a)-(b) $\mathbf{v}_{\perp}$ in the $x=0$ and $z=4$ planes respectively for the symmetric model. (c)-(d) $\mathbf{v}_{\perp}$ in the same planes for the asymmetric case with $m=0.5$. The contours and arrows denote $\eta|\mathbf{J}|$ and $\mathbf{v}_{\perp}$ respectively. The spine is shown in blue as a line in the $x=0$ plane and a square in the $z=4$ plane. The fan plane is shown in red. The parameters are as in Figure \ref{fig:fan-etaj}.}
\label{fig:vel-m0}
\end{figure}

\subsection{The Symmetric Case}
Before considering the asymmetric model, the symmetric one is first developed as a reference. In the symmetric case, closed form solutions can be achieved through the choice of $g(z)=1$ giving
\begin{equation}
Y=y-\frac{2j}{3}\left(\frac{2}{7}\frac{z^2}{h^2}M\left(1,\frac{11}{4},\frac{z^2}{h^2}\right)+1\right)ze^{-\frac{z^2}{h^2}-\frac{(zx^2)^2}{l^6}},
\end{equation}
where $M(\mu,\nu,x)$ is the \emph{Kummer M} hypergeometric function and
\begin{equation}
\Phi_{ni}=-\frac{j\eta_{0}B_{0}}{\mu_{0}L_{0}}\left[A+B+C\right]xe^{-\frac{z^2}{h^2}-\frac{z^{2}(x^2+Y^2)}{l^6}},
\end{equation}
where
\begin{eqnarray}
A&=&\left(1-\frac{2}{l^6}(zx^2)^2\right)(2B+1), \\
B&=&\frac{8}{21}\frac{z^4}{h^4}M\left(1,\frac{11}{4},\frac{z^2}{h^2}\right)+\frac{2}{3}\frac{z^2}{h^2},\\
C&=&\frac{8}{5 l^6}z^{4}x^{2}M\left(1,\frac{9}{4},\frac{z^2}{h^2}\right).
\end{eqnarray}
Figure \ref{fig:fan-etaj}(a)-(b) shows the localization of the non-ideal region and current flow and Figure \ref{fig:vel-m0}(a)-(b) the induced perpendicular plasma velocity. Due to its shape, the non-ideal region only affects a finite amount of magnetic flux and once the magnetic field threads into the non-ideal region it never exits. Therefore, this model can be considered to be the kinematic equivalent of the fan reconnection solutions of Craig and coworkers \cite{CraigFabling1996,Heerikhuisen2004} (where \emph{all} the flux enters the non-ideal region) modified to include current/resistivity localization perpendicular/parallel to the direction of shear. 

The result of localizing the influence of the non-ideal region is to induce cyclic flows within the envelope of field lines which just touch the edge of the non-ideal region. Such cyclic flows have also been seen before in the context of pure kinematic solutions for `finite-B' reconnection \cite{Hornig2003}. In particular, the flows in the finite-B case have an opposite vorticity either side of the non-ideal region. Despite the different field geometry, we see something similar appearing here with multiple vortices aligned to the spine lines.

\subsection{The Asymmetric Case}
To introduce asymmetry into the model we now choose
\begin{equation} 
g(z)=1+m\,\erf\left(\frac{z}{p}\right),
\label{asymsp-g}
\end{equation}
where $\erf(x)$ is the error function and $0\leq m \leq 1$. When $m=0$ or $p\to \infty$, $g(z)=1$ and the symmetric analytical solution above is recovered. On the other hand, when $m=1$ and $p\to 0$, $g(z)$ is double the heavyside (unit step) function and the magnetic field perturbation is switched off where $z<0$. Thus, a simple measure of the degree of system asymmetry is given by the factor $m/p$.

This choice of $g(z)$ allows for an analytical closed form solution for $K_{-1}(z)$ given by
\begin{equation}
K_{-1}(z)= \left[\frac{4}{5}tz^{2} M\left(1,\frac{9}{4},t z^{2}\right)+ 1 \right]\frac{4m}{\sqrt{\pi}p} e^{-tz^{2}},
\end{equation}
where $t=(h^{2}+p^{2})/h^{2}p^{2}$. However, no closed form solutions exist for $I_{a}(z)$ in this case so these integrals are found numerically by casting the general integral in the form
\begin{equation}
\frac{d I_{a}(z)}{dz} = \frac{1}{2z}I_{a}(z) + z^{\frac{a+1}{2}} e^{-\frac{z^{2}}{h^{2}}} g(z), \label{difsystem}
\end{equation}
ignoring the homogeneous solution and using a fourth order accurate Runge-Kutta scheme with the value of each integral at $z=0$ (given by the values of the symmetric solutions) used as the initial value. 

\begin{SCfigure*}
\centering
\includegraphics{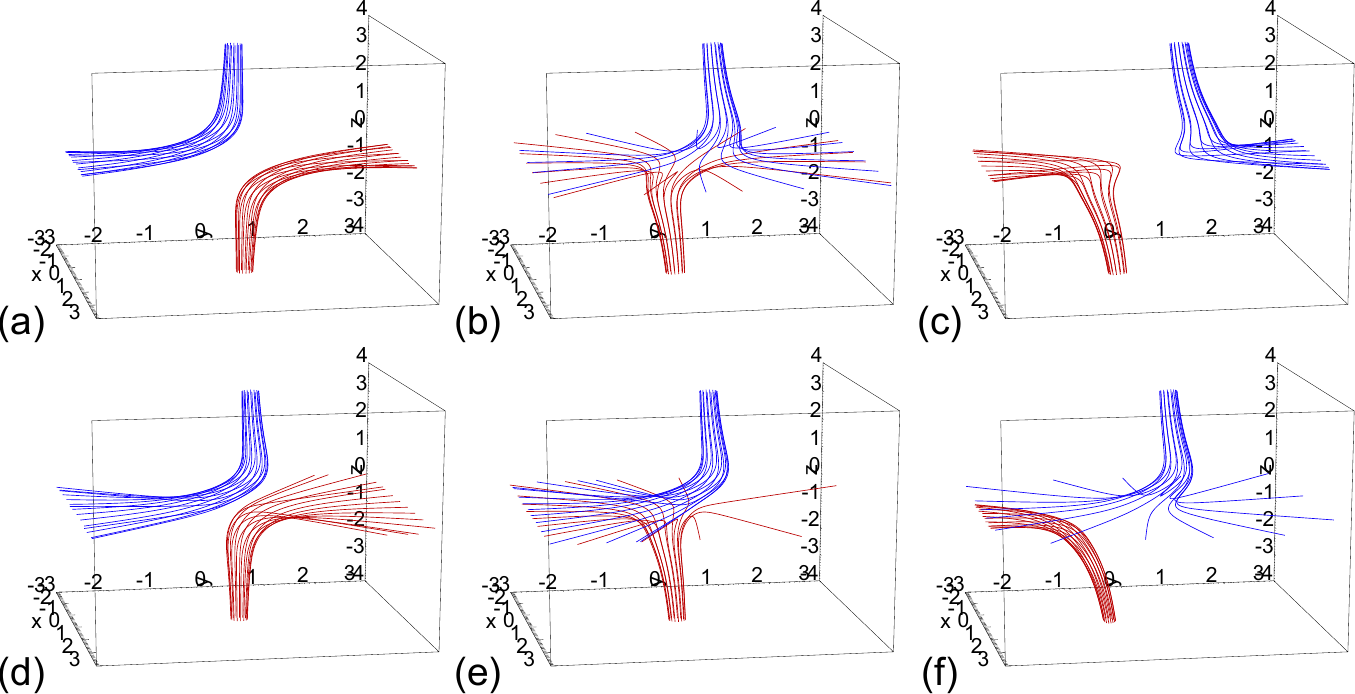}
\caption{Evolution of flux in the symmetric (a)-(c) and asymmetric (d)-(f) fan cases. For the parameter set given in Figure \ref{fig:fan-etaj}.}
\label{fig:flxtubes-fan}
\end{SCfigure*}

The skewed shape the non-ideal region and current flow now take is shown in Figure \ref{fig:fan-etaj}(c)-(d). We now denote the region where $z>0$ as the strong shear region and where $z<0$ the weak shear region. As a consequence of the weakened perturbation in the weak shear region the current flow is reduced there, with the converse being true of the region of strong shear. However, the current at the null remains the same as the symmetric case. It is evident from Figure \ref{fig:vel-m0}(c)-(d) that the induced plasma flow is strongly affected by the asymmetry. With the flow in the weak shear region dominating in strength over the strong shear flow. Most strikingly it is clear that the stronger flows of the weak shear region have crossed over the fan plane and flows over the top of the null. Evaluating $\mathbf{v}_{\perp}$ at the null yields
\begin{equation}
\mathbf{v}_{\perp}(0,0,0)=\left(0, -\frac{2j\eta_{0}}{\mu_{0}\sqrt{\pi}}\frac{m}{p},0\right).
\end{equation} 
Thus, for asymmetric fan reconnection a bulk flow of plasma occurs across the null point. This flow is a function of the degree of asymmetry of the system ($m/p$). How this affects the manner of connection change across the spine lines can be seen by tracking the movement of flux tubes bound to fluid elements in the ideal regions near each spine for the two cases (Figure \ref{fig:flxtubes-fan}). The presence of asymmetry clearly leads to different rates of flux transfer past each spine. This seems to be a generic feature of asymmetric fan reconnection brought on by the even nature of the $K_{-1}(z)$ function. 

In summary, asymmetric fan reconnection is characterized by asymmetric flux transfer past the spine lines and a non-zero plasma flow across the null which depends on the degree of asymmetry in the sheet.

\section{Asymmetric Spine Reconnection}
\label{sec:4} 
To create spine reconnection solutions it is more convenient to work in cylindrical coordinates so that now
\begin{equation}
\mathbf{B}_{n}=\frac{B_{0}}{L_{0}}\left(r,0,-2z\right).
\end{equation}
To this a perturbation function localized in $r$, of the following form, is added:
\begin{eqnarray}
\mathbf{B}_{p}&=&F(r,\phi) \; \mathbf{\hat{z}}\nonumber \\
&=&\frac{jB_{0}}{L_{0}} f(\phi) r e^{-\frac{r^{2}}{h(\phi)^{2}}} \; \mathbf{\hat{z}}.
\end{eqnarray}
The field line equations are then given by 
\begin{equation}
r=r_{0} e^{B_{0}s/L_{0}}, \quad \phi=\phi_{0}, \quad Z=Z_{0} e^{-2B_{0}s/L_{0}},
\end{equation}
where $Z_{0}(r_{0},\phi_{0},z_{0})$ is a constant of integration and 
\begin{eqnarray}
Z&=&z-e^{-2B_{0}s/L_{0}}\int \! e^{2B_{0}s/L_{0}} F(r,\phi) \, \mathrm{d}s \nonumber \\
&=&z+jf(\phi)\frac{h(\phi)^2}{r^{2}}D \label{Z}.
\end{eqnarray}
with
\[D=\frac{r}{2} e^{-\frac{r^2}{h(\phi)^2}}-\frac{h(\phi)}{4}\sqrt{\pi}\erf\left(\frac{r}{h(\phi)}\right). \]
In this case, flux surfaces are defined by $C_{1}(Zr^2)=const.$ and $C_{2}(\phi)=const.$, where $C_{1}$ and $C_{2}$ are arbitrary functions. A resistivity is then chosen of the form
\begin{equation}
\eta=\eta_{0}e^{-\frac{(Zr^2)^{2}}{k^6}}, \label{etasp}
\end{equation}
to complete the localization of the non-ideal ($\eta |\mathbf{J}|$) term. Using Equation (\ref{Phi}) the resulting electric potential is given by
\begin{equation}
\Phi_{ni}=-\frac{j \eta B_{0}}{\mu_{0} L_{0}}\left[\frac{\sqrt{\pi}h(\phi)}{2} f^{'}(\phi) \erf\left(\frac{r}{h(\phi)}\right)-2D f(\phi)\frac{h^{'}(\phi)}{h(\phi)} \right].\\
\end{equation}

\begin{figure}
\centering
\includegraphics{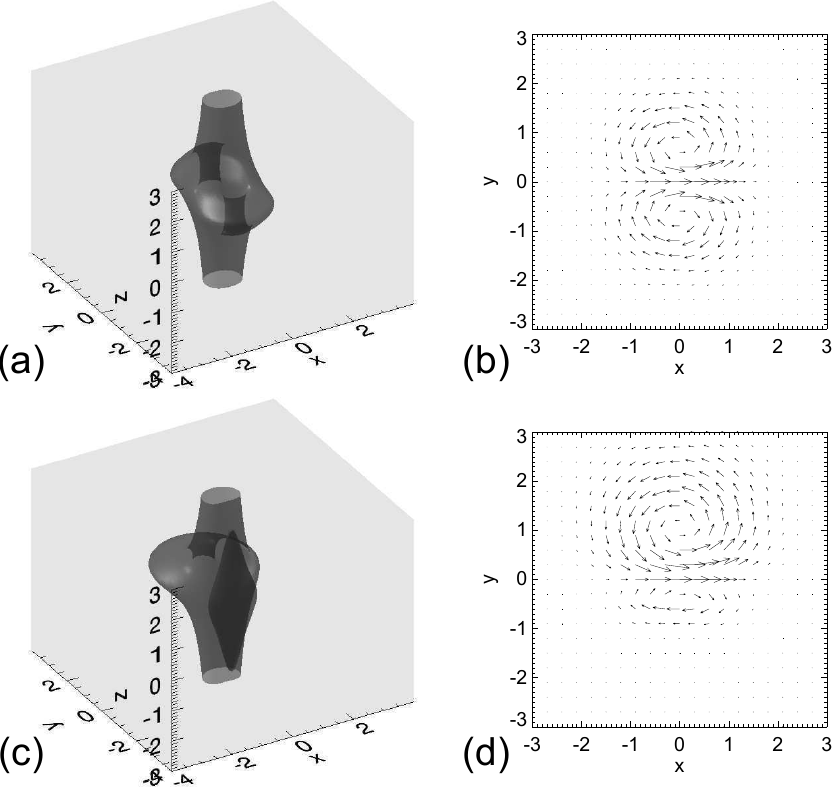}
\caption{(a)-(b) isosurface of $\eta|\mathbf{J}|$ at 25\% of the maximum and the current flow in the $z=0$ plane for the symmetric model. (c)-(d) the equivalent figures for the asymmetric model with $m=0.5$. Both have the parameter set $(B_{0},L_{0},\eta_{0},\mu_{0},j,L,k)=(1,1,1,1,2,1,1)$.}
\label{fig:spine-etaj}
\end{figure}

\subsection{The Symmetric Case}
Let us start by modeling the symmetric case which will be used as a benchmark for comparison once asymmetry is introduced. In particular, symmetric spine reconnection may be modeled by the choice of 
\[f(\phi)=\sin(\phi) \quad \& \quad h(\phi)=L.\]
This model can again be considered to be a kinematic extension of those of Craig and coworkers \cite{CraigFabling1996,Heerikhuisen2004}, with the spine aligned non-ideal region containing a finite magnetic flux due to the localization of $\eta$ along the spine. 

\begin{figure}
\centering
\includegraphics{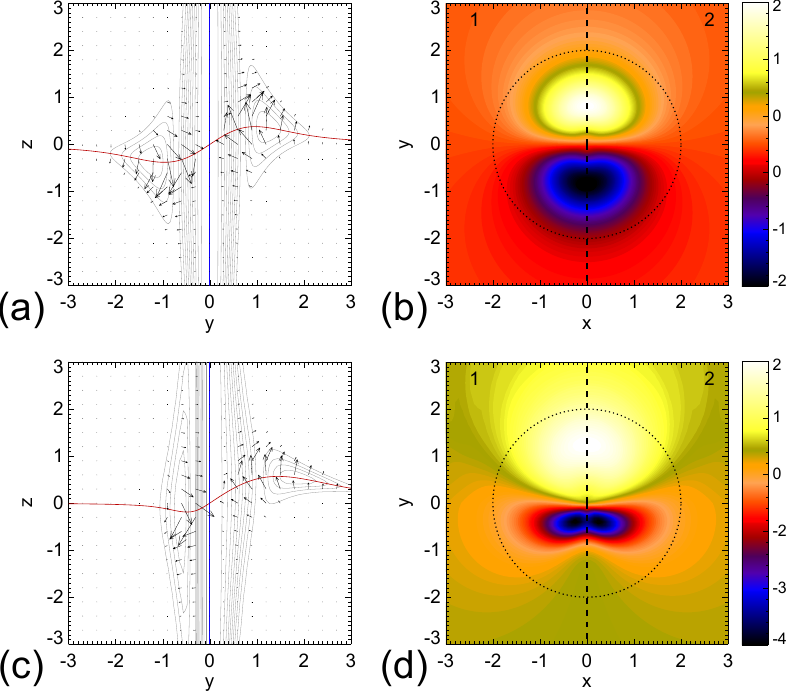}
\caption{(a) $\mathbf{v}_{\perp}$ in the $x=0$ plane with contours showing the strength of $\eta|\mathbf{J}|$. The spine is in blue and the fan plane red. (b) $v_{\perp z}$ evaluated on the fan plane ($Z=0$) with the dotted circle showing the cut taken in Figure \ref{fig:spine-velphi}. (c)-(d) the corresponding figures for the simple asymmetric case. For the parameters given in Figure \ref{fig:spine-etaj}.}
\label{fig:spine-vel}
\end{figure}

\begin{figure}
\centering
\includegraphics{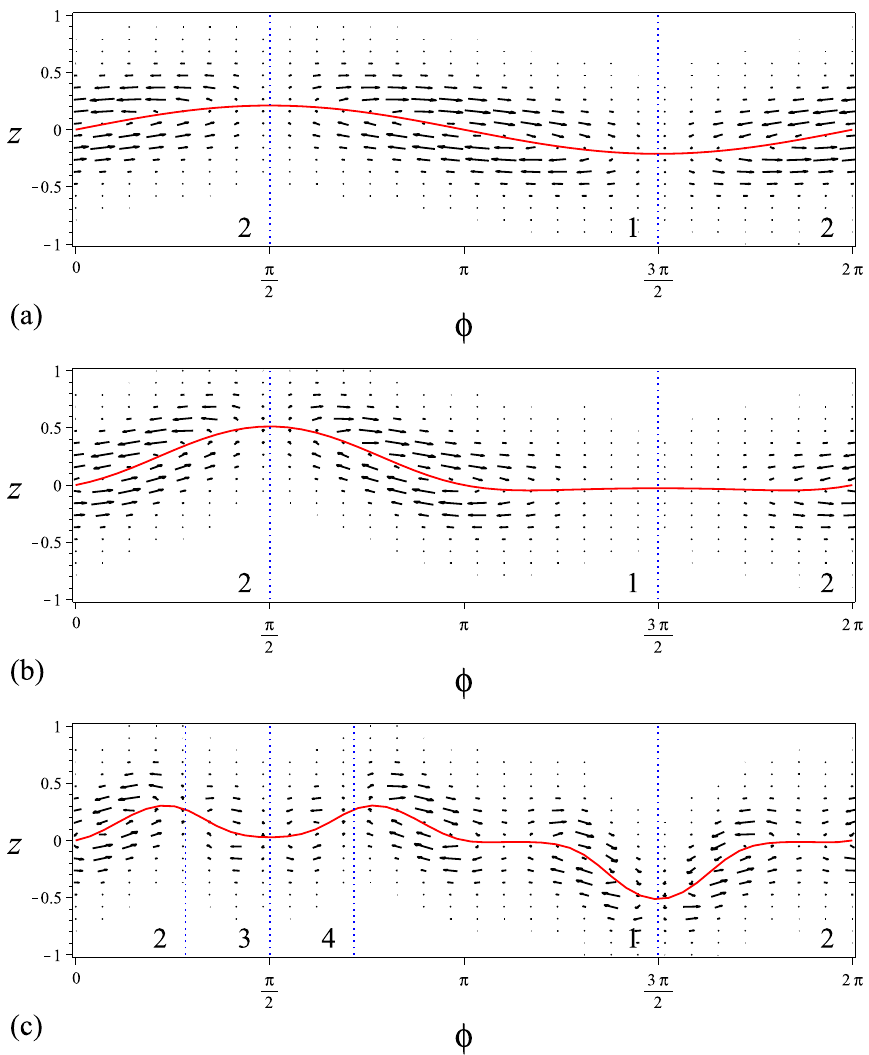}
\caption{$\mathbf{v}_{\perp}(r=2)$ with (a) $n=0$, (b) $n=1$ and (c) $n=3$. To be compared against Figures \ref{fig:spine-vel}(b), (d) and \ref{fig:vel-sp-n3}(b) respectively. For the parameter set $(B_{0},L_{0},\eta_{0},\mu_{0},j,L,k,m)=(1,1,1,1,2,1,1,0.5)$.}
\label{fig:spine-velphi}
\end{figure}

Figure \ref{fig:spine-etaj} shows how the non-ideal and current flow regions align to the spine axis. Since this is a pure solution, we would again expect cyclic flows within the envelope of flux which just touches the edge of the non-ideal region. Figures \ref{fig:spine-vel}(a)-(b) and \ref{fig:spine-velphi}(a) show how the resulting plasma flows reconnect flux equally and oppositely across the fan plane, rotate it around the spines and then return it back across the fan again. In other words, flux is continually rotated about the line $y=z=0$. This is most easily seen in the $z\phi$-plane as two vortices of opposite vorticity centered on $\phi=0$ and $\pi$. This creates two distinct regions (denoted 1 and 2) within which magnetic flux moves back and forth. 

\begin{SCfigure*}
\centering
\includegraphics{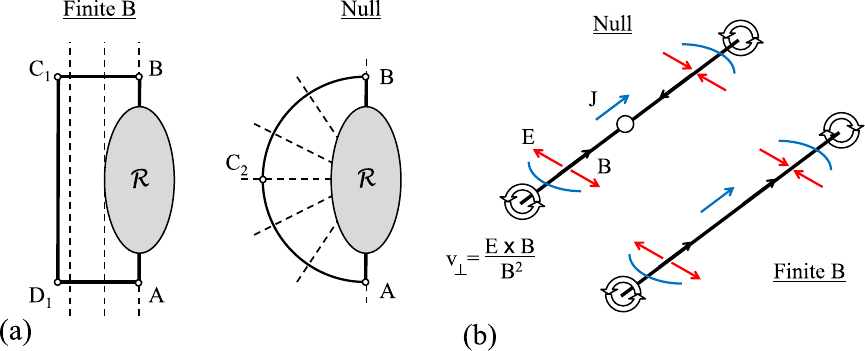}
\caption{(a) Integral loops constructed along paths either $\perp$ or $\parallel$ to the magnetic field. Such paths enable potential drops ($\Phi_{B}=\int{E_{\parallel} dl}+\Phi_{A}$) along field lines crossing $\mathcal{R}$ to be compared with flux movement in the ideal region. (b) The induced flux transport in the ideal region threading $\mathcal{R}$ with the edge of $\mathcal{R}$ depicted by blue lines.}
\label{fig:rot-diagram}
\end{SCfigure*}

Such induced plasma flows are a direct result of the underlying field geometry, and are in fact linked to the counter flows shown to be a generic feature of isolated finite-B reconnection \cite{Hornig2003}. To understand this, consider the diagram in Figure \ref{fig:rot-diagram}(a). Since $\mathbf{E}\cdot\mathbf{B}\neq 0$ inside the non-ideal region in the finite-B and spine cases there is a potential difference (say a drop) between $A$ and $B$ so that at $A$, $\Phi=\Phi_{A}$ and at $B$, $\Phi=\Phi_{B}$ where $\Phi_{A} > \Phi_{B}$. Following both loops back to $A$ the potential must increase again from $\Phi_{B}$ to $\Phi_{A}$. In the finite-B case, along the path between $C_{1}$ and $D_{1}$ ($C_{1}D_{1}$) the potential is fixed since in the ideal region $\mathbf{E}\cdot\mathbf{B}=0$.

Since the electric field requires a directional derivative, this results in oppositely directed electric fields along $C_{1}B$ and $D_{1}A$. As the background magnetic field is constant this leads to counter rotational flows (Figure \ref{fig:rot-diagram}(b)). A similar argument applies to the case when there is a null, except that now the electric potential varies smoothly between $B$ back to $A$ in the ideal region so that the electric field is oppositely directed along $C_{2}B$ and $C_{2}A$. However, as the background field contains a null the field switches direction along the axis of rotation $AB$, matching the sign change in $\mathbf{E}$ so the end result is the co-rotating flow described above (Figure \ref{fig:rot-diagram}(b)). This argument remains true for any finite non-ideal region within which $E_{\parallel}\neq 0$ and so applies beyond resistive MHD.

\subsection{A Simple Asymmetric Case}
A simple and analytically tractable way to incorporate current sheet asymmetry into the previous solution is through $h(\phi)$. \citet{Wyper2012} noted that asymmetric driving pulses lead to asymmetric spine-fan current sheets shifted in the direction of shear. This scenario seems the most plausible in practice and so we begin by modeling this via the choices of  
\[ f(\phi)= \sin(\phi) \quad \& \quad h(\phi)=L(1+m\sin(\phi)),\]
where $0\le m\le 0.5$ so that when $m=0$ the symmetric case is recovered. In this model, the parameter $m$ provides a measure of the degree of current sheet asymmetry. The skewed shape this gives the non-ideal and current regions is shown in Figure \ref{fig:spine-etaj}(b)-(c) when $m=0.5$. The asymmetric current rings correspond to a strong wide deformation of the fan plane on one side and a weak narrow one on the other. Figure \ref{fig:spine-vel}(c)-(d) shows how the general form of the cyclic flow remains, with plasma flowing through the fan plane on one side, looping around the spine and passing back through on the other within two distinct regions ($1$ and $2$). However, now the axis of rotation has been shifted into the strong shear semi-plane (Figure \ref{fig:spine-velphi}(b)). The rate at which plasma passes through the fan plane in each is now different with plasma flow across the weak shear region increased relative to the symmetric case. The inverse is true of the strong shear region. So, like asymmetric fan reconnection, the strongest outflows occur on the weakly deformed side. Why is this the case? To answer this it is convenient to first introduce the reconnection rate for the system and discuss both together.

\begin{figure}
\centering
\includegraphics[width=6.75cm]{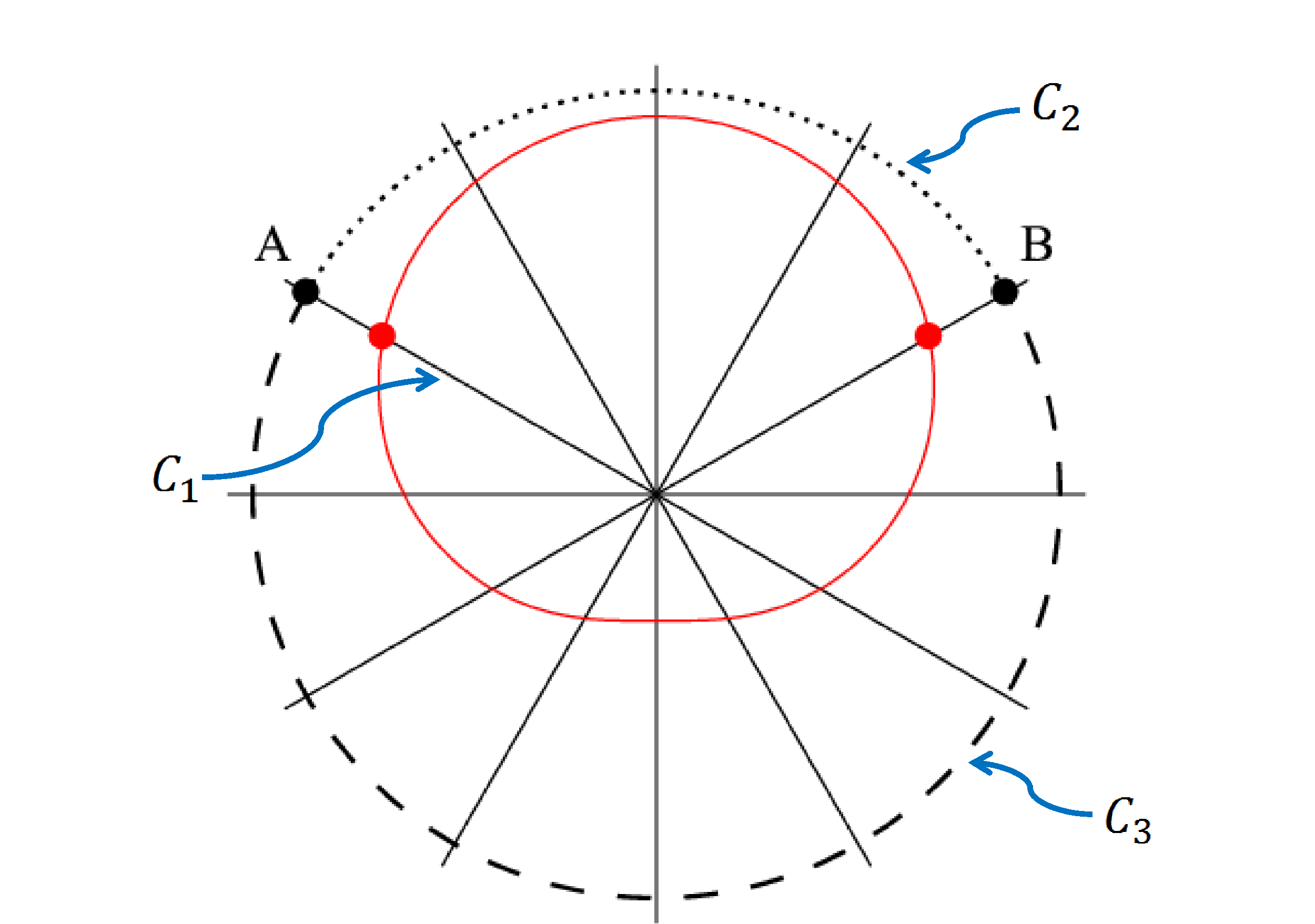}
\caption{Reconnection rate diagram. The edge of a general asymmetric non-ideal region is shown in red on the fan plane. The points $A$ and $B$ lie between the positive and negative regions of flux transport across this plane. These points can be connected by a path through the ideal region around the edge of the large side of the non-ideal region ($C_{3}$), around a path circuiting the small side ($C_{2}$) or though the non-ideal region and the null ($C_{1}$).}
\label{fig:rr-asym}
\end{figure}

\section{Reconnection Rate: The Simple Asymmetric Case}
\label{sec:rrsimpleas}
In symmetric spine-fan reconnection the reconnection rate is associated only with the transfer of magnetic flux across the fan plane \citep{Pontin2005}. The flow of flux across this plane is due solely to the spine reconnection aspect of it. As such, the same methodology is applicable to pure spine reconnection. The rate of flux transfer in one direction across the fan plane, in the ideal region outside of the current sheet, is taken as the reconnection rate of flux in this direction through that plane. For the strong shear region this can be measured by
\begin{equation}
%\frac{d\Psi}{dt}
\dot{\Psi}=-\int_{C_{2}}{\mathbf{v}\times\mathbf{B}\cdot d\mathbf{l}},
\end{equation}
where $C_{2}$ is the path shown in Figure \ref{fig:rr-asym}. Since $\mathbf{E}=-\mathbf{v}\times\mathbf{B}$ in this region and the integral of the electric field is path independent (as $\mathbf{E}=-\boldsymbol{\nabla} \Phi$ is conservative) this can be found from
\begin{equation}
%\frac{d\Psi}{dt}
\dot{\Psi}=\int_{C_{1}}{E_{\parallel} dl}=|\Phi_{B}-\Phi_{A}|. \label{spinerr}
\end{equation}
Here $A$ and $B$ are points in the fan plane lying between the regions of positive and negative flux transfer (outside of the non-ideal region) about which plasma flows circulate and $C_{1}$ is the path along the radial field lines between them (see Figure \ref{fig:rr-asym}). This again shows the similarity to the finite-B scenario where the potential drop along the axis of rotation is also the measure of the reconnection rate \cite{Hornig2003}. Since in steady state the integral of electric field is path independent:
\begin{equation}
\int_{C_{2}}{\mathbf{v}\times\mathbf{B}\cdot d\mathbf{l}} = \int_{C_{3}}{\mathbf{v}\times\mathbf{B}\cdot d\mathbf{l}}.
\end{equation}
Thus, an equal and opposite amount of flux must be transferred across the fan plane by the weak shear region. This explains why in the smaller weak field region the plasma jet is more intense than in the wider strong shear one. The strong shear region has a wider area over which to spread the same flux transfer. Therefore, asymmetric spine reconnection, in contrast to the fan case, is inherently equal and opposite in how it reconnects flux.% across its topological feature (the fan plane). 

To measure the rate of flux transfer in both directions across the fan plane in this asymmetric model requires the asymptotic value of the non-ideal electric potential at large radii ($r\gg L$) evaluated on the fan plane ($Z=0$) given by
\begin{equation}
\Phi_{ni}(r\gg L) = -\frac{j\eta_{0} B_{0}\sqrt{\pi}}{2\mu_{0}L_{0}}\left( h(\phi) f^{'}(\phi) + h^{'}(\phi) f(\phi)\right). \label{rr}
\end{equation}
For these choices of $f(\phi)$ and $h(\phi)$ this becomes
\begin{equation}
\Phi_{ni}(r\gg L) = -\frac{j\eta_{0} B_{0}\sqrt{\pi}}{2\mu_{0}} \frac{L}{L_{0}} \left(1+2m\sin(\phi)\right)\cos{\phi} . \label{Phi-asymt}
\end{equation}
Depending on the value of $m$, this potential will change and therefore will give different reconnection rates. In particular, the reconnection rate in both directions across the fan plane is given by double the difference between the maximum and minimum of this function. These occur at $\phi_{max}=\phi_{1}$ \& $\phi_{min}=\pi-\phi_{1}$ respectively, where $\phi_{1}$ is the lowest positive solution of 
\begin{equation}
\sin{\phi_{1}}=-\frac{1}{8m}\pm\frac{1}{2}\sqrt{\frac{1}{16m^2}+2}.
\end{equation}
The reconnection rate is then given in terms of this angle as 
\begin{equation}
\dot{\Psi}=\frac{2j\eta_{0}B_{0}\sqrt{\pi}}{\mu_{0}}\frac{L}{L_{0}} \left(1+2m\sin(\phi_{1})\right) \cos(\phi_{1}),
\end{equation}
or expressing it in terms of the reconnection rate of the symmetric case
\begin{equation}
\dot{\Psi}=\dot{\Psi}_{m=0} \left(1+2m\sin(\phi_{1})\right) \cos(\phi_{1}).
\end{equation}
Therefore, it is found that in the simplest asymmetric scenario the reconnection rate changes depending upon the degree of asymmetry, but the manner of flux transport across the fan plane remains an equal and opposite process. 

Lastly, it could be argued that the above result may be a consequence of the steady state condition. However, consider the integral of the electric field around some closed path $\mathcal{C}$ ($C_{2}+C_{3}$ in Figure \ref{fig:rr-asym}) enclosing the entire non-ideal region in the fan plane in the general time dependent case. Then
\begin{eqnarray}
\int_{\mathcal{C}}{\mathbf{E}\cdot d\mathbf{l}}&=&\int_{\mathcal{C}}{\nabla \times\mathbf{E}\cdot d\mathbf{S}} \nonumber\\
&=&-\int_{\mathcal{C}}{\frac{\partial \mathbf{B}}{\partial t} \cdot d\mathbf{S}} =0,
\end{eqnarray}
where $\mathbf{S}$ is a surface on fan plane enclosed by the closed curve $\mathcal{C}$ for which $\mathbf{B}\cdot \mathbf{S}=0$ by definition. Thus, on the fan plane
\begin{equation}
\int_{\mathcal{C}}{\mathbf{v}\times\mathbf{B}\cdot d\mathbf{l}}=0.
\end{equation}
Therefore, even in time dependent systems the reconnection of flux across the fan plane (in contrast to reconnection across the spine lines) is always an equal and opposite process. Note also, this argument relies only upon there being a localized non-ideal region in the fan plane for which $E_{\parallel}\neq0$ and not on the non-ideal term itself or its extent along the spine. Therefore, this argument applies in general beyond the confines of resistive MHD and to spine-fan reconnection.

\begin{SCfigure*}
\centering
\includegraphics{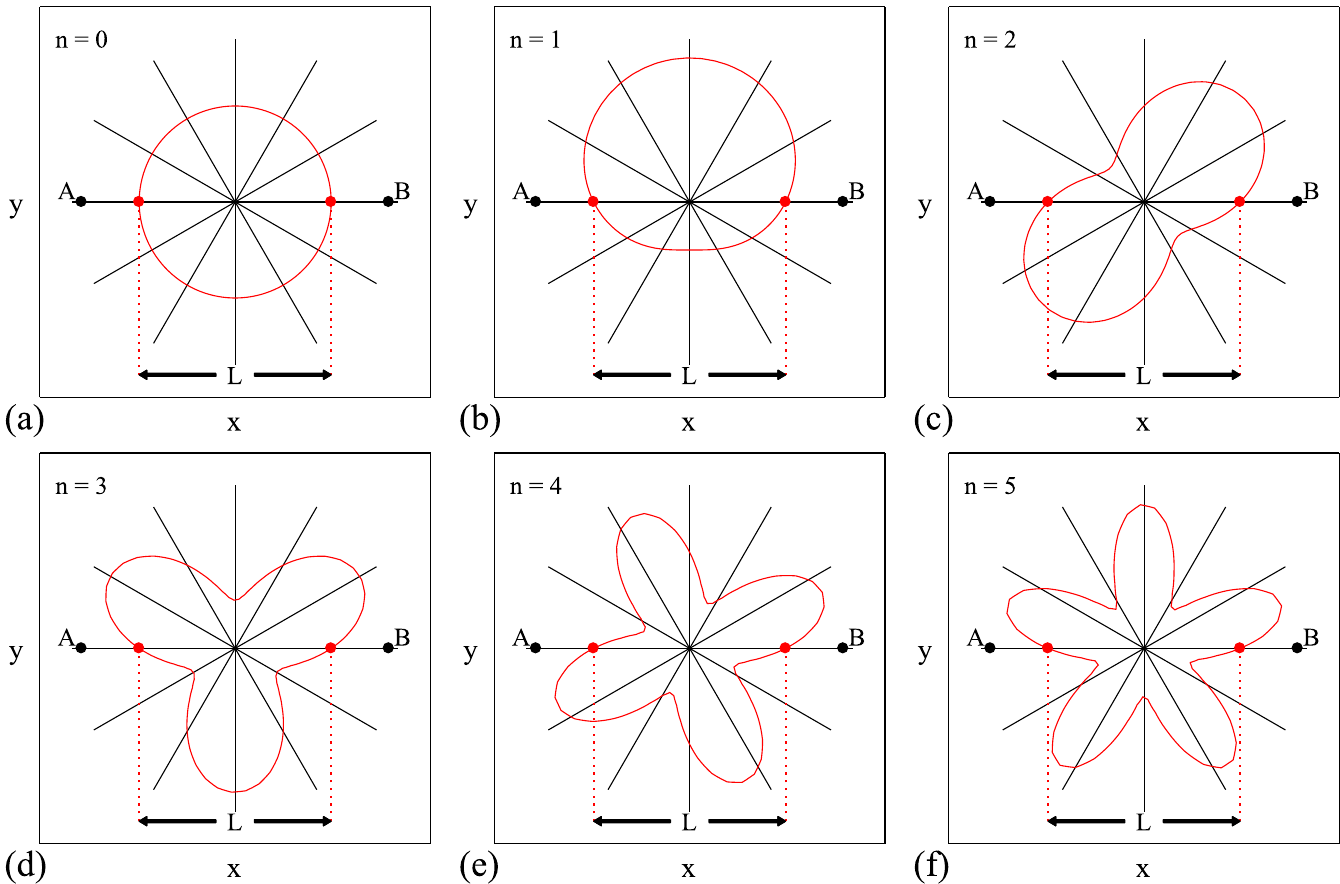}
\caption{The shape of the non-ideal region shown in red on the fan plane when $h(\phi)=L(1+m\sin{n\phi})$. The distance $L$ indicates the length of the non-ideal region along the line $AB$.}
\label{fig:spine-sin-modes}
\end{SCfigure*}

\begin{figure}
\centering
\includegraphics{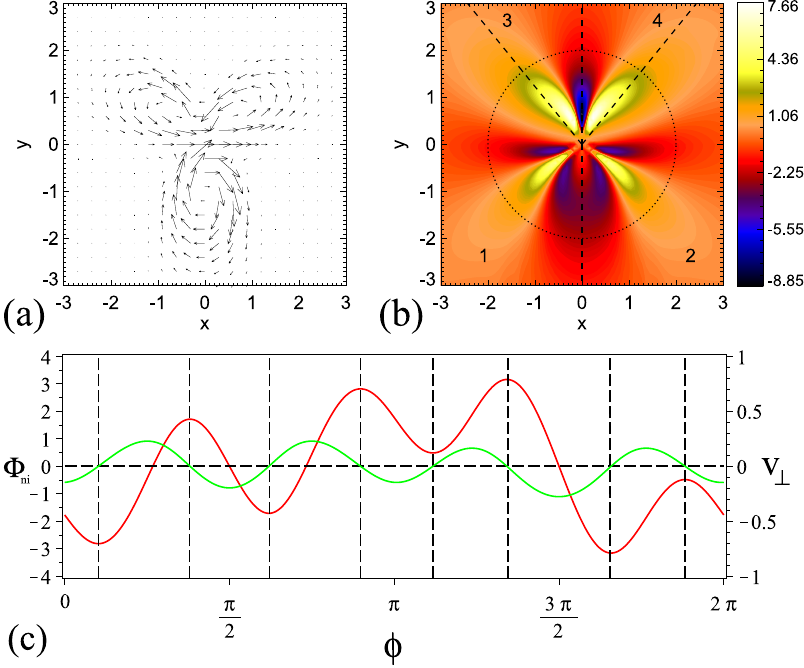}
\caption{Asymmetric spine reconnection with $n=3$. (a) Current flow in the $z=0$ plane. (b) $v_{\perp z}$ in the fan plane ($Z=0$) to be compared against Figure \ref{fig:spine-velphi}(c). (c) $\Phi_{ni}(r=4,Z=0)$ (red) overlayed with $\mathbf{v}_{\perp}(r=4,Z=0)$ (green). The overlaid dashed grid highlights the relationship between the two quantities.}
\label{fig:vel-sp-n3}
\end{figure}

\section{Asymmetric Spine Reconnection: General Examples} 
\label{sec:asymgen}
Let us now consider more complex examples of asymmetric spine reconnection and generalize some of the ideas presented in the previous section. The above are in fact part of a family of solutions given by 
\[ f(\phi)=\sin(\phi) \quad h(\phi)=L(1+m\sin(n\phi)). \]
Figure \ref{fig:spine-sin-modes} shows the projection of the non-ideal region on to the fan plane for the first five modes. To be clear, each of these modes has the \emph{same} current flow at the null, but the shape of the current sheet varies as $n$ is changed. The value of $n$ dictates how many lobe-like extensions of the non-ideal region there are and so, along with $m$, is a measure of the sheet deformation. For $n\geq 2$ (and $m>0.5$ when $n=1$, where small lobes also appear) these lobe-like extensions produce plasma flow back and forth within each semi-plane and can considerably complicate the plasma dynamics near the fan.

Figures \ref{fig:vel-sp-n3}(a)-(b) and \ref{fig:spine-velphi}(c) show current and plasma flows when $n=3$. As with all the models flux is cycled continuously, however, now distinct regions of contained flux movement (regions $3$ and $4$) occur. Within regions $3$ and $4$ a single vortex cycles magnetic flux around continuously, reconnecting it back and forth across the fan plane. In regions $1$ and $2$ a similar large vortex flow is present, but within in it two internal vortices coexist with a stagnation point between them. Depending upon where flux initially starts in regions $1$ and $2$ it will find itself either trapped to circulate around within one of these internal vortices or around the edges of both. Regions $1$ and $2$ are roughly speaking analogous to the two regions discussed in the previous sections when $n=1$ and $2$ as flux is, in general, brought through the fan plane in the positive direction in the $y>0$ semi-plane and sent back through the fan in the negative direction in the $y<0$ semi-plane. Regions $3$ and $4$ have no direct counter part as flux is trapped to circulate within the wedge defined by them. For modes with larger $n$ the number of these flux transport regions and the number of vortices internal to them (like the two vortices in region $1$ for instance) increase. 

These additional wedges and internal vortices make the definition and interpretation of the rate at which flux is reconnected across the fan plane more difficult. On the one hand, the total physical amount of flux reconnected across the fan is given by the sum of all flux cycled back and forth by every vortex flow (including those internal to each flux transport region). This quantity gives the genuine reconnection rate of the system. On the other hand, the wedges of contained flux transport and internal vortex flows that do not straddle the line $y=z=0$ give a net zero contribution to the flux driven through the semi-plane they lie in. If the plasma flows of the non-ideal region are assumed to be coupled to the global environment through an ideal stagnation point flow of the traditional type, i.e. one that brings in flux to be reconnected in equal and opposite directions across each semi-plane, then such internal vortices do not contribute to the global rate at which flux is `seen' to cross the fan plane by the global field. The net transfer of flux through one semi-plane is given, in this case, by the potential difference along the line $y=z=0$ ($AB$ in Figure \ref{fig:spine-sin-modes}).

Therefore, in general, for spine reconnection two reconnection rates can be defined. A local reconnection rate given by the sum of the potential drops between adjacent maxima and minima in the electric potential, evaluated in the fan plane outside of the non-ideal region ($r\gg L$ and $Z=0$). These maxima and minima correspond to points in the fan plane where $\mathbf{v}_{\perp}=0$, so lie either in the centers of the vortices or the stagnation points between two vortices of like vorticity. As such, this potential drop gives the total flux transfer between these zero points. Figure \ref{fig:vel-sp-n3}(c) shows this relationship between $\Phi_{ni.}$ and $\mathbf{v}_{\perp}$ evaluated in the ideal region ($r\gg L$) on the fan plane ($Z=0$) for $n=3$. Since around the entire non-ideal region the flux transfer is equal and opposite, this quantity can be expressed as double the sum of the difference between each maxima and the next minima ahead of it in $\phi$:
\begin{equation}
\dot{\Psi}_{local}(N)=2\sum_{k=1}^{N}{|\Phi_{max}(\phi_{k})-\Phi_{min}(\phi>\phi_{k})|},
\end{equation} 
where $N$ is the total number of peaks in the electric potential. Alternatively, a global reconnection rate can be defined which gives the net flux transfer through both semi-planes:
\begin{equation}
 \dot{\Psi}_{global}=2|\Phi(\phi=0)-\Phi(\phi=\pi)|,
\end{equation}
quantifying the rate that an observer far from the null `sees' flux reconnected at the null if the ideal flow is of a stagnation point type. The definitions of each then lead to the following properties when all other parameters are fixed:
\begin{eqnarray}
\dot{\Psi}_{local}&=&\dot{\Psi}_{global}; \; n = 0, \\
\dot{\Psi}_{local} & > & \dot{\Psi}_{global}; \; n \geq 1, \\
\dot{\Psi}_{local}(N+1) &\geq & \dot{\Psi}_{local}(N).
\end{eqnarray}
Thus, this local rate will always at least equal that of the global rate. Under this new definition the reconnection rate found in the simple asymmetric case ($n=1$) in Section \ref{sec:rrsimpleas} becomes the local rate. Modes with high $n$ can be used to describe the situation when the edge of the current sheet is deformed by the action of an instability such as the tearing mode or Kelvin-Helmholtz instability. During such deformations this local rate would be expected to dwarf the global one.

For these choices of $f(\phi)$ and $h(\phi)$ the global rate is given by 
\begin{equation}
\dot{\Psi}_{global}=\frac{2j\eta_{0}B_{0}\sqrt{\pi}}{\mu_{0}} \frac{L}{L_{0}},
\end{equation}
which is independent of both the degree of asymmetry ($m$) and the number of lobes ($n$) as a result of the fact that the length of the non-ideal region along the line $AB$ always remains fixed as $L$ (Figure \ref{fig:spine-sin-modes}). Thus, even in the situation when the edge of the sheet is fragmented (and if it is the net transfer that is of interest) then it is the length scale along the line $AB$ that dictates the global reconnection rate. When this length scale is not conserved by the manner of sheet asymmetry the global reconnection rate is simply dictated by this changing length scale ($L_{n}$):
\begin{equation}
\dot{\Psi}_{global}=\frac{2j\eta_{0}B_{0}\sqrt{\pi}}{\mu_{0}} \frac{L_{n}}{L_{0}}.
\end{equation}
For example, the choice of 
\[f(\phi)=\sin(\phi), \quad h(\phi)=L(1+m\cos(n\phi))\]
leads to a changing length scale of $L_{n}=L \left(1+m(-1)^n\right)$ depending on whether two lobes are in or out of phase along the line $AB$. The global reconnection rate in this case, therefore, has two distinct rates.

\subsection{Reconnection Rate vs Ohmic Dissipation}
\begin{SCfigure*}
\centering

\includegraphics{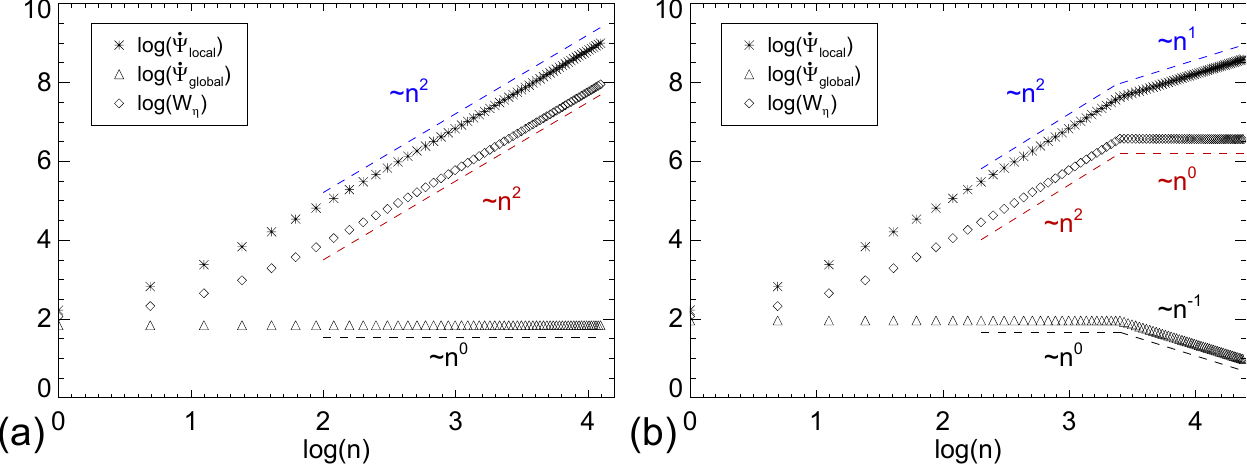}
\caption{Log-log plots of $\dot{\Psi}_{local}$, $\dot{\Psi}_{global}$ and $W_{\eta}$ vs $n$. (a) when all other parameters are held fixed (given in Figure \ref{fig:spine-velphi}). (b) when a stall is introduced heuristically into $W_{\eta}$ (see Equation \ref{jn}).}
\label{fig:n-scaling}
\end{SCfigure*}

Lastly, we now consider how these two reconnection rates compare against ohmic heating. For simplicity, consider the case when $\eta=\eta_{0}$ so that the non-ideal region is invariant along the direction aligned to the spine. The ohmic dissipation per unit length in this direction is then given by
\[W_{\eta}=\int{\eta J^{2} dV}= \eta_{0}\int{J(r,\theta)^{2} r dr d\theta} \]
and the local and global reconnection rates are still as given above. How each quantity varies as the mode parameter $n$ (and so the sheet deformation) increases is shown in Figure \ref{fig:n-scaling}(a). $\dot{\Psi}_{global}$ remains fixed, as discussed above, but for large values of $n$ both $W_{\eta}$ and $\dot{\Psi}_{local}$ increase toward the same power law dependence $\sim n^{2}$. This shows that I) the above axioms relating $\dot{\Psi}_{global}$ to $\dot{\Psi}_{local}$ are upheld and II) when all other parameters are fixed it is the complete transfer of flux across the fan plane ($\dot{\Psi}_{local}$) not the rate at which an observer far from the null sees it cross ($\dot{\Psi}_{global}$) which is related to ohmic dissipation. However, such a decoupling of $\dot{\Psi}_{global}$ from $W_{\eta}$ seems unlikely as in practice if a current sheet is fragmented following an instability (essentially changing to a sheet with higher $n$) the current density will not increase indefinitely but is likely to stall. It might be more reasonable to expect that the ohmic dissipation within the layer would increase only up to a point. Such a stall can be included heuristically by introducing a dependence of $j$ upon $n$ such that
\begin{equation}
j(n)= 
\begin{cases}
j_{0}, & n \le n_{0} \\
\frac{j_{0}n_{0}}{n}, & n > n_{0}
\end{cases}
\label{jn}
\end{equation}
Figure \ref{fig:n-scaling}(b) shows the transition in the scalings of each when $n_{0}=30$. Once $W_{\eta}$ stalls, the global transfer of flux across the fan plane reduces at the rate at which the local rate increases ($\sim n$) as energy available to globally reconnect flux is now expended in cyclic local flux transfer. 

In a self consistent system it is likely that, following say some instability which filaments the current layer, both effects will be observed. That is, there will be both an increase in ohmic dissipation/local reconnection rate \emph{and} a decrease in the global rate that flux is transferred across the layer. This has been observed recently in numerical simulations of a similar situation when the torsional fan current sheet is fragmented via the Kelvin-Helmholtz instability \cite{Wyper2013}.

\section{Concluding Remarks}
\label{sec:conclusion}
This paper has been concerned with investigating the role of current sheet asymmetry in the reconnection modes of spine and fan reconnection. Through a series of analytical models we have shown that the behavior of each is rather different. Asymmetric fan reconnection is characterized by an asymmetric reconnection of flux past each spine line and a bulk flow of plasma across the null point. A behavior masked in previous 3D symmetric studies \cite{Pontin2005,CraigFabling1996,PriestTitov1996}, but that shares some general characteristics with the asymmetric 2D study of \citet{Cassak2007} in that asymmetric magnetic fields in the inflow regions lead to a non-zero flow of plasma across the null. In contrast to asymmetric fan reconnection, asymmetric spine reconnection is inherently equal and opposite in how flux is reconnected across the fan plane. However, with an extra degree of freedom, asymmetric spine reconnection is considerably more complex. In the simplest asymmetric case, asymmetric outflow jets were formed within the vicinity of the null but globally an equal and opposite magnetic flux is driven through both sides of the fan plane. Higher modes were also found where constrained regions of flux transport were localized to within wedges in each semi-plane (Figure \ref{fig:vel-sp-n3}). In this more complex situation, two definitions for reconnection rate became appropriate: a local reconnection rate quantifying how much flux is genuinely reconnected across the fan plane and, on the assumption that the non-ideal region has been created through some background ideal stagnation point flow, a global rate associated with the net flux driven across each semi-plane. Such a two part definition has already been shown to be useful in the similar scenario of torsional fan reconnection subject to current sheet fragmentation following the KH instability \cite{Wyper2013}. The choice of background ideal flow used to advect flux into and away from the non-ideal region is crucial for the interpretation of the reconnection rate. Therefore, different composite solutions could potentially give rise to different reconnection rates depending upon how much of the flux transfer within each vortex flow can be accessed by the global ideal flow field. An investigation of the composite solutions would be interesting to pursue in the future.

These models also provide a link between the exact incompressible solutions with current sheets of reduced dimensionality \cite{CraigFabling1996,Heerikhuisen2004} traditionally used to investigate spine and fan reconnection and the localized kinematic solutions of spine-fan \cite{Pontin2005,AlHachimi2010} and finite-B \cite{Hornig2003} reconnection utilizing localized resistivities in fields of constant current flow. In particular, the spine scenario shows how both null point and finite-B reconnection are driven by fundamentally the same process ($\mathbf{E}\cdot\mathbf{B}\neq0$ leading to a potential difference) but with a different knock on effect in terms of flux transport facilitated by the local magnetic field structure. This also ties in nicely with the separator model of \citet{Wilmot-Smith2011}. These authors developed a time dependent kinematic model for reconnection occurring along the separator joining two magnetic nulls. They found that singular (since in their model the field near the nulls is ideal) cyclic flows were driven at the nulls by reconnection along a single separator and that as the non-ideal region grows stronger multiple separators form. The cyclic flows they observed are described well by our symmetric spine model and one could postulate that the wedges of constrained transport found in cases with a highly deformed current sheet may be linked to where multiple separators rejoin the null. This may also be interesting to pursue in future.

\begin{acknowledgments}
P.F.W. and R.J. wish to acknowledge the financial support of EPSRC.
\end{acknowledgments}

\bibliographystyle{aipnum4-1}
%\bibliography{biblionew}


\end{document}